\documentclass[doublecol]{epl2} 

\usepackage{graphicx}
\usepackage{dcolumn}
\usepackage{bm}
\usepackage{hyperref}
\usepackage{amssymb}
\usepackage{mathrsfs}
\usepackage{amsmath,fleqn}
\newcommand{\bea}{\begin{eqnarray}}
\newcommand{\eea}{\end{eqnarray}}
\newcommand{\ba}{\begin{array}}
\newcommand{\ea}{\end{array}}

\newcommand{\be}{\begin{equation}}

\newcommand{\ee}{\end{equation}}
\newcommand{\bt}{\begin{teo}}
\newcommand{\et}{\end{teo}}

\newcommand{\la}{\lambda}

\title{The intermediate level statistics in dynamically localized chaotic eigenstates}
\shorttitle{Dynamical localization in chaotic systems} 

\author{B. Batisti\'c \inst{1} \and T. Manos\inst{1,2} \and M. Robnik\inst{1}}
\shortauthor{B. Batisti\'c \etal}

\institute{
  \inst{1} CAMTP - Center for Applied Mathematics and Theoretical Physics, University of Maribor, Krekova 2, SI-2000 Maribor, Slovenia, European Union.\\
  \inst{2} School of Applied Sciences, University of Nova Gorica, Vipavska 11c, SI-5270 Ajdov\v s\v cina, Slovenia, European Union.
}


\abstract{We demonstrate that the energy or quasienergy level spacing
distribution in dynamically localized chaotic eigenstates is excellently
described by the Brody distribution, displaying the fractional power
law level repulsion. This we show in two paradigmatic systems,
namely for the fully chaotic eigenstates
of the kicked rotator at $K=7$, and for the chaotic eigenstates
in the mixed-type billiard system (Robnik 1983), after separating the
regular and chaotic eigenstates by means of the Poincar\'e Husimi function,
at very high energies with great statistical significance (587654 eigenstates,
starting at about 1.000.000 above the ground state).
This separation confirms the Berry-Robnik picture, and is performed
for the first time at such high energies.}

\begin{document}

\maketitle

\section{Introduction} \label{Intro}

One of the main findings in quantum chaos \cite{Stoe,Haake,Rob1998}
of stationary Schr\"odinger
equation is the fact that the statistics of spectral fluctuations
of the discrete quantal energy spectra around the smooth mean behaviour
of classically chaotic systems obeys the Random Matrix Theory (RMT),
in terms of the Gaussian ensembles of random  matrices
\cite{Stoe,Haake,Mehta,GMW}, in the sufficiently
deep (or far) semiclassical limit. This  finding is known as the Bohigas -
Giannoni - Schmit (BGS) conjecture first published in \cite{BGS},
although some preliminary ideas were introduced in \cite{Cas}.
By ``sufficiently deep (or far) semiclassical
limit" we mean that some semiclassical condition is satisfied,
namely that all relevant classical transport times, like the typical
ergodic time, or diffusion time, are smaller than the so-called Heisenberg
time, or break time, given by $t_H=2\pi\hbar/\Delta E$, where $h=2\pi\hbar$
is the Planck constant and $\Delta E$ is the mean energy level spacing,
such that the mean energy level density is $\rho(E) =1/\Delta E$.
If the stated condition is satisfied, the quantum eigenstates as
represented in the ``quantum phase space" by the Wigner functions, or
Husimi functions, are uniformly extended \cite{Rob1998}, and the spectral
statistics is as in RMT, namely like for Gaussian Orthogonal Ensemble (GOE)
or Gaussian Unitary Ensemble (GUE), depending on the antiunitary symmetries of
the system \cite{Stoe,Haake,Rob1998,Mehta,RB1986,Rob1986}. Here we  treat only the GOE case. As $t_H \propto \hbar^{1-N}$, where $N$ is the number of degrees of freedom ($N\ge 2$), it is clear that at sufficiently small $\hbar$ the Heisenberg time $t_H$ becomes larger than any classical transport time. This is the ultimate semiclassical (deep or far) regime.

Chaotic eigenstates are either in classically fully chaotic (ergodic)
systems or in mixed-type systems, where regular and chaotic eigenstates
are supported by the classically regular and chaotic regions coexisting in the phase space, respectively. In the mixed-type case the Berry-Robnik scenario \cite{BR1984} applies, resting on the so-called Principle of Uniform Semiclassical Condensation of Wigner functions on classical invariant components \cite{Rob1998}.

However, quite generally, if the semiclassical condition is not satisfied, such that $t_H$ is no longer larger than the relevant classical transport time, like e.g. the diffusion time in fully chaotic but slowly ergodic systems, we find the so-called dynamical localization, or Chirikov localization.
Dynamical localization was first discovered in time dependent systems \cite{CCFI79}.  It was intensely studied since then in particular by Chirikov, Casati, Izrailev, Shepelyanski and Guarneri, in the case of the kicked rotator as reviewed in \cite{Izr1990}. See also \cite{Izr1988,Izr1986,Izr1987,Izr1989,MR2013}. For a review of the Floquet systems see \cite{Stoe, Haake}. It has been observed that in parallel with the localization of the eigenstates one finds the fractional power law level repulsion (of the quasienergies) even in fully chaotic regime (of the finite dimensional kicked rotator), and it is believed that this picture applies also to time independent (autonomous) Hamilton systems and their eigenstates \cite{MR2013,Pro2000}. Indeed, this has been analyzed indirectly with
unprecedented precision and statistical significance recently by Batisti\'c and Robnik \cite{BatRob2010} in case of mixed-type systems,
and is confirmed directly in the present work.

The present work concerns for the first time the theoretical separation of the regular and chaotic eigenstates, assumed to exist in the Berry-Robnik picture \cite{BR1984}, and the analysis of the chaotic eigenstates and the corresponding energy subspectrum, using the Poincar\'e Husimi functions.  The complete details will be published in a separate paper \cite{BatRob2013}. An early attempt of separation of eigenstates in the billiard system has been published in \cite{LiRob1995}, using a quite different approach at much lower energies and with much smaller statistical significance. Another study \cite{BB2007} in mushroom billiards concerned the eigenstates at much lower energies and with much smaller statistical significance, where the
dynamical tunneling effects were investigated but not the dynamical localization effects, although a clear
deviation from the GOE behaviour in chaotic eigenstates was found.

Here we show that, after the separation of regular and chaotic eigenstates, the level spacing distribution of the chaotic part of the spectrum obeys the Brody distribution with a very high precision and statistical significance, and a similar observation is found in the spectrum of quasienergies of the finite dimensional kicked rotator in the classically fully chaotic regime. Thus, the Brody distribution captures correctly the spectral fluctuation properties of dynamically localized chaotic eigenstates, which is the first such a clear demonstration.

The Brody distribution \cite{Bro1973,Bro1981} is
\be \label{BrodyP} \nonumber
P_B(S) = C_1 S^{\beta} \exp \left( - C_2 S^{\beta +1} \right),
\ee
where $C_1$ and $C_2$ are determined by $<1> = <S> = 1$,
\be \label{Brodyab}
C_1 = (\beta +1 ) C_2, \;\;\; C_2 = \left( \Gamma \left( \frac{\beta +2}{\beta +1}
 \right) \right)^{\beta +1},
\ee
with  $\Gamma (x)$ being the Gamma function. The Brody parameter $\beta$ is in the  interval $[0,1]$, where $\beta=0$ yields the Poisson distribution in case of the strongest localization, and $\beta=1$ gives the Wigner surmise (2D GOE, as an excellent approximation of the infinite dimensional GOE), which describes the extended chaotic eigenstates. It turns out that the Brody distribution fits the empirical data much better than e.g. the distribution function proposed by F. Izrailev (see \cite{Izr1988,Izr1989,Izr1990} and the references therein), defined as
\begin{flalign} \label{IzrailevDistrib}
      P_I(S) = A\left (\frac{1}{2}\pi S \right )^{\beta} \exp \left [-\frac{1}{16}\beta \pi^2 S^2 -\left (B-\frac{1}{4}\pi \beta \right )S \right],
\end{flalign}
where the constants $A$ and $B$ are determined by the normalizations $<1>=<S>=1$.

If in a mixed-type system the couplings between the regular eigenstates and chaotic eigenstates become important, at low energies, due to the dynamical tunneling, we can use the ensembles of random matrices that capture these
effects \cite{VSRKHG,GroRob1,GroRob2,BatRob2010}. As the tunneling strengths typically decrease exponentially with the inverse effective Planck constant,
they rapidly disappear with increasing energy, or by decreasing the value of the Planck constant. In this work we shall deal only with very high-lying eigenstates in billiards, and therefore we can neglect the effects of tunneling, whilst
in the kicked rotator all eigenstates are chaotic and thus there is no tunneling at all.

\section{Billiard} \label{Billiard}

The  billiard domain ${\cal B}$ under study is defined, as introduced in \cite{Rob1983,Rob1984}, by the quadratic conformal map of the unit circle $|z|=1$ of the $z$-complex plane onto the boundary $\partial{\cal B}$ in the $w$-complex plane (which is the physical plane) as
$w= z + \la z^2$.
We choose $\la=0.15$, in which case the boundary is convex and the dynamics is of the mixed-type. There are regions of regular motion near the boundary (Lazutkin's caustics) and also in the interior \cite{Laz1981,Laz1991,Hayli1987}. Classical dynamics is fully determined by the bounce map on the phase space cylinder $(s,p)$, where the arclength parameter $s\in [0,{\cal L})$ determines the position of the collision point on the boundary, whilst $p\in [-1,+1]$ is the sine of the reflection angle and is thus the momentum conjugate to $s$. In fact, due to the reflection symmetry and the time-reversal symmetry, one quadrant $(s,p)\in [0,{\cal L}/2] \times [0,1]$
is enough to be presented. By $\rho_1$ we denote the relative phase space
volume (not to be confused with the area on the Poincar\'e surface of section) of the classical regular components and by $\rho_2=1-\rho_1$ its complement, the relative volume of the chaotic component. In our case $\rho_1=0.175$ \cite{BatRob2010}.

The quantum mechanics is determined by the stationary Schr\"odinger equation which in appropriate units is just the Helmholtz equation for the billiard domain ${\cal B}$,
\be \label{Helm}
\Delta \psi + k^2 \psi =0,
\ee
with the Dirichlet boundary condition $\psi=0$ on the boundary $\partial{\cal B}$. The energy eigenvalues $E_j=k_j^2$, $j=1,2,3,\dots$ define the energy spectrum of the billiard, and after spectral unfolding, based on the Weyl formula with perimeter corrections (see e.g. \cite{Stoe}), we study its statistical properties. The total spectrum can be conceptually decomposed into regular and chaotic eigenstates. The further we are in the semiclassical regime, the ``cleaner" is this separation. The physical separation is possible by looking at the structure of the Wigner functions \cite{Wig1932}, or better, Husimi functions \cite{Hus1940} (which are Gaussian averaged Wigner functions). In doing this for a great number of eigenstates we first introduce quantum
representation analogous to the classical one. To this end we remind that the {\em boundary function} $u(s)$, which is the normal derivative of $\psi ({\bf r})$ on the boundary at the point $s$,
$u(s) = {\bf n}\cdot \nabla_{{\bf r}} \psi \left({\bf r}(s)\right)$,
where ${\bf n}$ is the unit outward vector  normal to the boundary at position $s$, uniquely determines the solution $\psi ({\bf r})$ at any point in the
interior of ${\cal B}$, by the relation (see e.g. \cite{BerWil1984})
\be \label{Intfromu}
\psi_j({\bf r})  = - \oint dt\; u_j(t)\; G\left({\bf r},{\bf r}(t)\right).
\ee
Thus, in certain analogy to the classical mechanics, the quantum mechanics is completely described by the boundary functions $u_j(s)$.
$G({\bf r},{\bf r'}) = -\frac{i}{4} H_0^{(1)}(k|{\bf r}-{\bf r'}|)$, is the free particle Green function,
where $H_0^{(1)} (x)$ is the Hankel function.

Having set up this representation we proceed by defining the {\em Poincar\'e Husimi functions} (PHF), following the formalism as in \cite{BFS2004}, which in turn is based on \cite{CPC1993,TV1995,SVS1997,VS1995}, namely we define the manifestly $s$-periodic coherent states (with the period ${\cal L}$)
\begin{flalign} \label{cohsta} \nonumber
c_{(q,p),k} (s) = \sum_{m\in \mathbb{Z}}
\exp \{ i\,k\,p\,(s-q+m{\cal L})\} \times \\
\exp \left(-\frac{k}{2}(s-q+m{\cal L})^2\right).
\end{flalign}
They are concentrated at $(q,p) \in [0,{\cal L}] \times \mathbb{R}$. Here we have omitted all normalization factors, because in the end we shall normalize the PHF anyway. Then, using this, the PHF associated with the $j$-th eigenstate represented by the boundary function $u_j(s)$ with the eigenvalue $k=k_j$, is
\be \label{Husfun}
H_j(q,p) = \left| \int_{\partial {\cal B}} c_{(q,p),k_j} (s)\;
u_j(s)\; ds \right|^2 ,
\ee
which is positive definite by construction. In the semiclassical limit $j \rightarrow \infty$, and $k_j\rightarrow \infty$, we shall observe that the PHF is concentrated on the classical invariant regions, which can be an invariant torus, a chaotic component, or the entire Poincar\'e surface of section $(s,p)$ if the motion is ergodic.

\begin{figure*}
 \begin{center}
 \includegraphics[width=7.5cm]{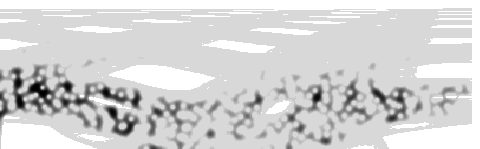}
 \includegraphics[width=7.5cm]{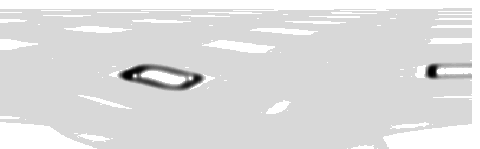}
 \caption{Examples of a chaotic (left) and a regular (right) state in the Poincar\'e Husimi representation. The corresponding parameter values are:
 $k_j= 2000.0181794,\; M_j=0.981$ and $k_j=2000.0777155,\; M_j= -0.821$. The gray background is the classically chaotic invariant component. We show only one quarter of the surface of section $(s,p)\in [0,{\cal L}/2] \times [0,1]$, because due to the reflection symmetry and time-reversal symmetry the four quadrants are equivalent.} \label{fig2}
 \end{center}
\end{figure*}

The quantum eigenstates were calculated using the method of Vergini and Saraceno \cite{VS1995} with great accuracy, for all eigenstates (587654) within the interval $k\in[2000,2500]$. The number of eigenstates below $k=2000$ is estimated by the Weyl rule as about 1.000.000. Then, when calculating
the PHF, the momentum $p$ is rescaled by the eigenvalue $k_j$, such that $p=1$ corresponds to the original $p=k_j$.  Two examples of the PHF are shown in figure \ref{fig2}. The relevant classical transport time for the global spreading on the chaotic component is estimated by numerical calculations as $N_T\approx 10^7$ collisions, and the semiclassical condition for the dynamical localization regime in our units is $k\le N_T/2$  \cite{BatRob2013}, which is very well satisfied. The tunneling effects disappear exponentially as $\propto \exp (- C k)$, where the constant $C>0$ is such that at $k \ge 2000$ the effects certainly entirely disappear as observed in \cite{BatRob2010}.

Next the PHF $H(q,p)$ (\ref{Husfun})  was calculated on the grid points ($400 \times 400$) on one quadrant $(s,p)\in [0,{\cal L}/2] \times [0,1]$ for each eigenstate $j$, then normalized such that the sum
over all grid points is equal to one,  and the overlap index $M$ was calculated according to the definition
\be \label{indexM}
M = \sum_{i,j} H_{i,j}\; A_{i,j}.
\ee
Here, $A_{i,j}$ is equal to $+1$ if the grid point $(i,j)$ belongs to the chaotic region, and $-1$ if it belongs to the regular region. Therefore, due to the normalization of $H_{i,j}$, and in the ideal (semiclassical) case $M$ is either $+1$ or $-1$. In practice, $M$ is not exactly $+1$ or $-1$, but can have a value in between. The reasons are two, first the finite discretization of the phase space (the finite size grid), and second,  the finite wavelength (not sufficiently small effective Planck constant, for which we can take just $1/k_j$). If so, the question is, where to cut the distribution of the $M$-values, at the threshold value $M_t$, such that all states with $M<M_t$ are declared regular and those with $M>M_t$ chaotic.
We introduce two natural criteria: {\bf (I)} {\em The classical criterion:} the threshold value $M_t$ is chosen such that we have exactly $\rho_1$
fraction of regular levels and $\rho_2=1-\rho_1$ of chaotic levels. {\bf (II)} {\em The quantum criterion:} we choose $M_t$ such that we get the best possible agreement of the chaotic level spacing distribution with the Brody distribution, which is expected to capture the dynamical localization effects
of the chaotic eigenstates.

\begin{figure*}
\center
\includegraphics[width=7.5cm]{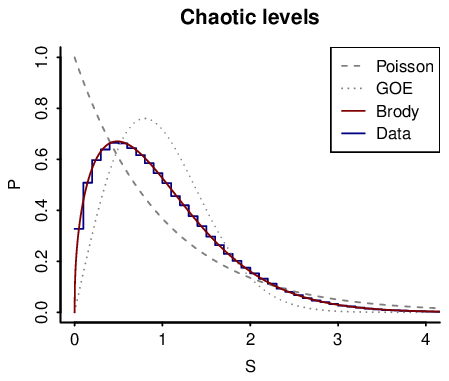}
\includegraphics[width=7.5cm]{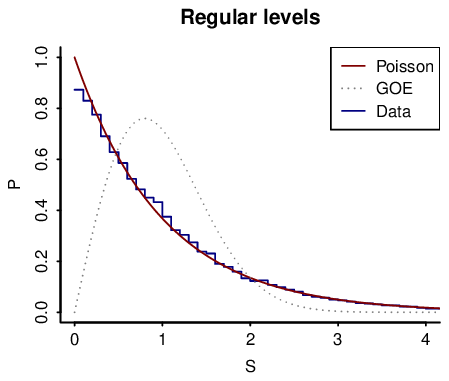}
\caption{ Separation of levels using the classical criterion $M_t=0.431$. {\bf Left}: The level spacing distribution for the chaotic subspectrum after unfolding, in perfect agreement with the Brody distribution $\beta=0.444$.
{\bf Right}: The level spacing distribution for the regular part of the spectrum, after unfolding, in excellent agreement with Poisson.} \label{separspectrum}
\end{figure*}

\begin{figure*}
\center
\includegraphics[width=7.5cm]{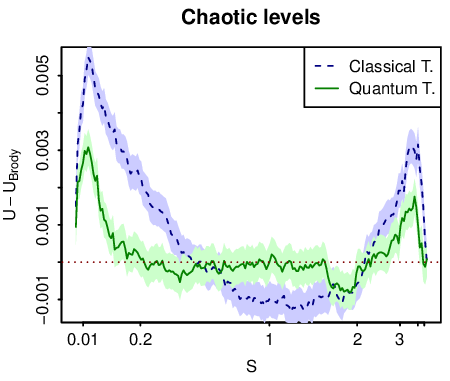}
\includegraphics[width=7.5cm]{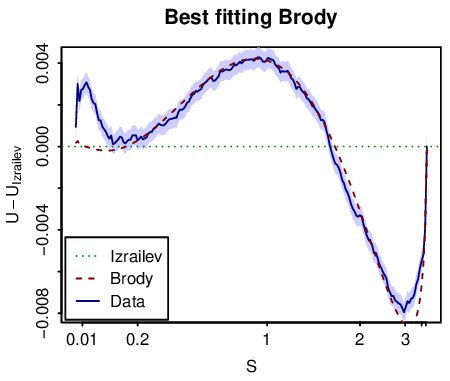}
\caption{{\bf Left}: The $U$-function plots as differences $U(data)-U(Brody)$ for the chaotic levels, for both criteria, the classical one $M_t=0.431$ and the quantum one $M_t=0.75$.
{\bf Right}:  We show the $U$-function plot for the chaotic levels, clearly  showing that Brody distribution (dashed) is much better, even perfect, description of the $U$-function than the Izrailev, namely in the case when Brody gives the
best possible fit,  it should be compared with Izrailev (on the abscissa) at the same quantum threshold $M_t=0.75$. The belts around  the data lines indicate the expected statistical $\pm$ one-sigma errors.}
\label{Uregcha}
\end{figure*}

In figure \ref{separspectrum} we show the level spacing distribution after separation using the classical criterion. Obviously, Brody distribution is
an excellent fit for chaotic levels with $\beta=0.444$. In figure \ref{Uregcha} we show the $U$-function plot (introduced in \cite{ProRob1994b}, and explained in detail in \cite{BatRob2010}), demonstrating in a more sensitive way that Brody distribution is indeed excellent description of the chaotic level spacings, in case of the criterion (II) even better than using the criterion (I).  We also show that the Izrailev level spacing distribution \cite{Izr1988,Izr1989,Izr1990} (\ref{IzrailevDistrib}) is much less significant than the Brody distribution. We have also demonstrated \cite{BatRob2013} that Brody is still a better fit even when we choose $M_t$ such that the Izrailev fitting is the best possible, and the same conclusion holds when $M_t$ is
chosen such that the Poisson fitting for the regular levels is the best possible.


\section{Kicked rotator} \label{Kicked rotator}

The kicked rotator introduced in \cite{CCFI79} is a paradigm of
Floquet systems in quantum chaos \cite{Izr1990}. The Hamiltonian is
\be  \label{KR}
H= \frac{p^2}{2I} + V_0 \,\delta_T(t)\,\cos \theta.
\ee
Here $p$ is the (angular) momentum, $I$ the moment of inertia, $V_0$ is the strength of the periodic kicking, $\theta$ is the (canonically conjugate,
rotation) angle, and $\delta_T(t)$ is the periodic Dirac delta function with period $T$. Since between the kicks the rotation is free, the Hamilton equations of motion can be immediately integrated,  and thus the dynamics can be reduced to the standard mapping,  or so-called Chirikov-Taylor mapping, given by
\be \label{SM2}
P_{n+1} = P_n + K \sin \theta_{n+1},\;\;\; \theta_{n+1} = \theta_n  + P_n,
\ee
where $P_n=p_nT/I$, and the system is now governed by a single classical {\em dimensionless} control parameter $K=V_0 T/I$, and the mapping is area preserving. The quantities $(\theta_n,p_n)$ refer to their values just
immediately after the $n$-th kick.

The quantum kicked rotator (QKR) is the quantized version of (\ref{KR}) (see \cite{Stoe}),
where now we have two {\em dimensionless} quantum control parameters
$k=V_0/\hbar, \;\;\; \tau= \hbar\; T/I$, which satisfy the relationship $K = k\tau = V_0 T/I$.  We have studied the case $K=5$, extensively investigated by Izrailev et al \cite{Izr1990},
where there is still a regular region in the phase space of relative area 2\%, and many more cases, $K=7$, where
the system is classically practically fully chaotic.
If $\tau/(4\pi)$ is a sufficiently irrational number the quasienergy spectrum is discrete.
 The infinite dimensional system exhibits in such cases the dynamical localization and consequently the Poisson level spacing distribution, for the quasienergies. However, if we study the finite dimensional system \cite{Izr1987,Izr1988,Izr1990}, which can be regarded as one of the possible discretizations of the kicked rotator (or of standard map), the scenario changes completely. Then it is convenient to work in the angular momentum basis, defined by $|n\rangle = \exp(i n \theta)$. We have studied the system
\begin{flalign} \label{u_repres}
u_n(t+T) = \sum_{m=1}^{N} U_{nm}u_m(t), \quad n,m=1,2,...,N.
\end{flalign}
The finite symmetric unitary matrix $U_{nm}$ determines the evolution of an $N$-dimensional vector, namely the Fourier transform $u_n(t)$ of $\psi(\theta,t)$, and is composed as follows
\be \label{Unm}
    U_{nm}=\sum_{n'm'}G_{nm'}B_{n'm'}G_{n'm},
\ee
where $G_{ll'}=\exp \left (i\tau l^2/4 \right )\delta_{ll'}$ is a diagonal matrix corresponding to free rotation during a half period $T/2$, and  the matrix $B_{n'm'}$ describing the one kick is
\begin{flalign} \label{Bnmoper}\nonumber
& B_{n'm'}= \frac{1}{2N+1}\times \\ \nonumber
& \sum_{l=1}^{2N+1} \left \{ \cos \left [ \left (n'-m' \right ) \frac{2 \pi l}{2N+1}\right ] - \cos \left [(n'+m')\frac{2 \pi l}{2N+1} \right ] \right \} \\
& \times  \exp \left [-ik\cos \left (\frac{2 \pi l}{2N+1}\right ) \right ].
\end{flalign}
The model (\ref{u_repres}-\ref{Bnmoper}) with a finite number of states is considered as the quantum analogue of the classical standard mapping on the torus with closed momentum $p$ and phase $\theta$, where $U_{nm}$ describes only the odd states of the systems, i.e. $\psi(\theta)=-\psi(-\theta)$, provided we have the case of the quantum resonance, namely $\tau =4\pi r/(2N+1)$, where $r$ is a positive integer. The matrix (\ref{Bnmoper}) is obtained by starting the derivation from the odd-parity basis of $\sin(n\theta)$ rather than the general angular momentum basis $\exp(in\theta)$.

Nevertheless, we use this model for any value of $\tau$ and $k$, as a model which in the resonant and in the generic case (irrational $\tau/(4\pi)$) corresponds to the classical kicked rotator, and in the limit $N\rightarrow \infty$ approaches the infinite dimensional model, restricted to the
symmetry class of the odd eigenfunctions. It is of course just one of the possible discrete approximations to the continuous infinite dimensional
model.

\begin{figure}
\center
\includegraphics[width=7.5cm]{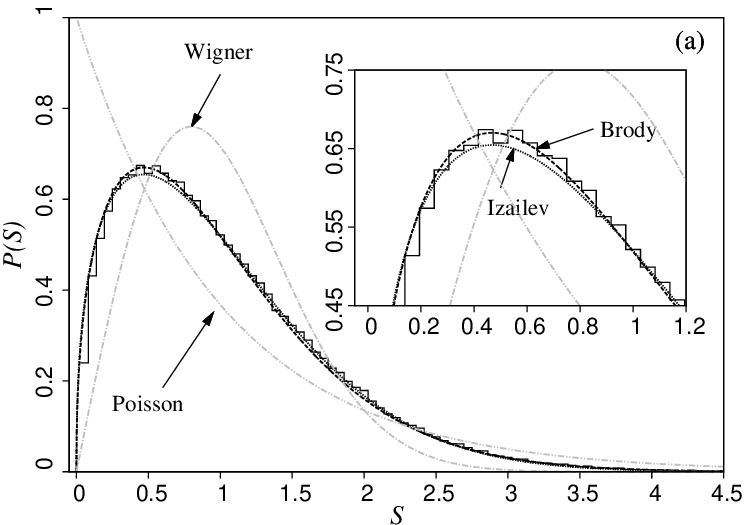}
\includegraphics[width=7.5cm]{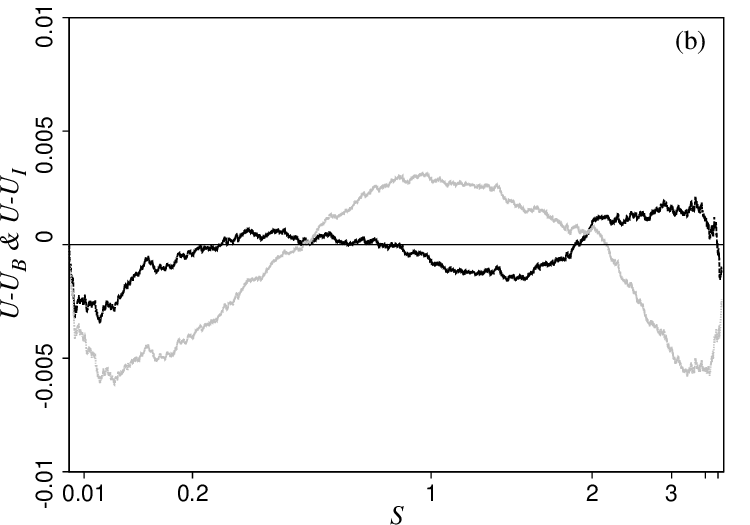}\\
\caption{(a): $P(S)$ (histogram - black solid line) of the model [Eqs.~(\ref{u_repres})-(\ref{Unm})] fitted with distribution $P_B(S)$ (black dashed line) and $P_I(S)$ (black dotted line) for 641 matrices of size 398, with  $K=7$ and $k$  in small interval around $11$. The gray lines indicate  Poisson and Wigner. In (b)
we show the difference of the numerical data and the best fitting Brody (black line) and Izrailev (gray line) PDFs by using the $U$-function. Thus in case of the ideal fitting the data would lie on the abscissa. In this case, based on the $P(S)$ fit we get $\beta=0.421$ for Brody and $\beta=0.416$ for Izrailev fit. }
\label{fig6}
\end{figure}

In figure \ref{fig6} we show the main results. We clearly see that the Brody distribution is an excellent fit to the level spacing distribution, even more clearly in the $U$-function plots, and we also see that it is empirically better than Izrailev's fit. We have also verified that the same conclusion is reached if we use the improved Izrailev distribution \cite{CIM1991}. The details will be published in a separate paper \cite{MR2013}, where we find that Brody distribution is better than Izrailev for all $\beta$, at $K\ge 7$.

\section{Conclusions} \label{Conclusion}

We have demonstrated that in case of dynamical localization of chaotic eigenstates the corresponding energy spectrum obeys very accurately the Brody distribution, better than the Izrailev distribution, and in both cases we observe the fractional power law level repulsion with the exponent $\beta\in [0,1]$. This has been demonstrated for the finite dimensional kicked rotator and its quasienergy spectra, as well as for the chaotic eigenstates in a time-independent Hamilton system, namely the convex billiard of the mixed-type
(Robnik 1983, $\la=0.15$; \cite{Rob1983,Rob1984}), and the corresponding energy levels. The separation of the regular and chaotic eigenstates has been performed  by means of Poincar\'e Husimi functions of the boundary function. The successful separation of course also confirms the Berry-Robnik picture \cite{BR1984} of separating the regular and chaotic levels in the semiclassical limit, where the tunneling effects can be neglected. Many more details  will appear in two separate papers \cite{MR2013} and \cite{BatRob2013}.

The theoretical derivation of the Brody level spacing distribution for the dynamically localized chaotic eigenstates  is still an open problem for the future. The billiard systems are suitable also for the experimental applications, like e.g. in quantum dots, and microwave cavities introduced and studied by H.-J. St\"ockmann \cite{Stoe}. We also propose to study from the present point of view the hydrogen atom in strong magnetic field as an example of classical and quantum chaos par excellence, as introduced in \cite{Rob1981,Rob1982,HRW1989,WF1989}, although the technical efforts  are much bigger than in billiard systems, where we have a great number of elegant numerical techniques \cite{VPR2007}, all of them used in \cite{BatRob2010}.

\acknowledgments
Financial support of the Slovenian Research Agency ARRS under the grants P1-0306 and J1-4004 is gratefully acknowledged.

\end{document}